\documentclass[preprint,superscriptaddress,nofootinbib,aps]{revtex4}
 \usepackage{graphicx}
 \usepackage{bm}
 \begin{document}

 %%%%%%%%%%%%  title  %%%%%%%%%%%%%%%%%%%%%%%%%
 \title{$B \rightarrow K K$ decays with the soft-gluon corrections}
 \thanks{This work is in part supported by the National Science
         Foundation of China Under Grant No 10075053}
 \author{Li Lin}
 \email[E-mail address: ]{lilin2009@mail.ihep.ac.cn}
 \affiliation{Institute of High Energy Physics,
              Chinese Academy of Sciences,\\
              P.O.Box 918(4),
              Beijing 100039, China}
 \author{Wu Xiang-Yao}
 \affiliation{Institute of Physics, NanKai University,
              Tianjin, 300071}
 \author{Huang Tao}
 \affiliation{Institute of High Energy Physics,
              Chinese Academy of Sciences,\\
              P.O.Box 918(4),
              Beijing 100039, China}
 \date{\today}

 %%%%%%%%%%%%  abstract %%%%%%%%%%%%%%%%%%%%%%%%%
 \begin{abstract}
 We analyze the $B \rightarrow KK$ decays with the soft-gluon
 corrections by using the QCD light-cone sum rules (LCSR).
 Although QCD factorization approach calculates the leading
 order factorization parts and the radiative corrections from
 hard- gluon exchanges at $\alpha_{s}$ order, it is worthwhile
 to estimate the nonfactorizable soft-gluon contributions from
 all the tree and penguin diagrams systematically. Our results
 show that the soft-gluon effects always decrease the branching
 ratios and give a few percentage corrections at most in the
 $B \rightarrow KK$ decays.

 \end{abstract}

 \maketitle

\section{INTRODUCTION}

Recently, A. Khodjamirian \cite{s1} has presented an approach to
calculate the hadronic matrix elements of nonleptonic B meson
decays within the framework of the light-cone sum rules, where
the nonfactorizable soft contributions can be effectively dealt
with. As we know, QCD factorization approach \cite{s2}
 provided that the hadronic
matrix elements for $B\rightarrow \pi \pi, K \pi$ decays
 can be expanded
in the
powers of $\alpha_{s}$
and $\frac{\Lambda_{QCD}}{m_{b}}$ and exhibited a considerablly strong
predicative potential. However, this approach can't calculate
$\frac{\Lambda_{QCD}}{m_{b}}$ corrections
quantitatively, such as the
nonfactorizable contributions from the soft-gluon exchanges. Thus
it is interesting to evaluate the corrections from the soft-gluon
exchanges by using light-cone QCD sum rules.

In the previous paper \cite{s3}, the role of the soft-gluon exchanges in
 $B\rightarrow \pi \pi$ has been studied by using the light-cone
QCD sum rules. Compared to the Ref. \cite{s1}, the calculations are
carried
out not only for the tree operators but also for the penguin ones.
Ref.\cite{s3} showed that the $\frac{\Lambda_{QCD}}{m_{b}}$ corrections
 from the soft-gluon exchanges are not
always negligible in the process
$B\rightarrow \pi \pi$ and the nonfactorizable soft contributions
are almost as important as the $O(\alpha_s)$ correction parts,
 and in some cases even have the same
order effects as that of the factorization amplitude. Therefore
it is worthwhile to evaluate the nonfactorizable soft-gluon
contributions in the process $B\rightarrow KK$.

\section{CORRELATOR AND SUM RULES}

Similar to the case of $B\rightarrow \pi \pi$, we can calculate the
contributions from the soft-gluon
exchanges in  $B \rightarrow KK$ including the tree and penguin
operators. We begin with the effective Hamiltonian $H_{eff}$ which is
responsible for the $B\rightarrow KK$ decays \cite{s4}:

\begin{eqnarray}
{\cal H}_{eff}=&&\frac{G_{F}}{\sqrt{2}}[V_{ub}V_{uq}^*(c_{1}(\mu)O_{1}(\mu)
+c_{2}(\mu)O_{2}(\mu)) \nonumber \\&&
%+V_{cb}V_{cq}^*(c_{1}(\mu)O_{1}^c(\mu)+c_{2}(\mu)O_{2}^c(\mu))
%\nonumber \\&&
-V_{tb}V_{tq}^*\sum \limits_{i=3}^{10} c_{i}O_{i}]+h_{.}c_{.}
\end{eqnarray}
where $O_{1,2}$ are the tree operators and $O_{3}-O_{10}$ denote the
penguin ones. By applying the Fierz transformation, the operators
which is related
 to the soft corrections to $B \rightarrow KK$ can be clearly presented
in effective weak Hamiltonian:

\begin{eqnarray}
{\cal H}_{eff}=\frac{G_{F}}{\sqrt{2}}[V_{ub}V_{ud}^*(c_{1}(\mu)+
\frac{c_{2}(\mu)}
{3})O_{1}(\mu)+2c_{2}(\mu)\widetilde{O}_{1}(\mu)+\cdots],
\end{eqnarray}
where
\begin{eqnarray}
{O}_{1}=(\overline{s}\Gamma_{\mu}u)
(\overline{u}\Gamma^{\mu}b),
\end{eqnarray}
and
\begin{eqnarray}
\widetilde{O}_{1}=(\overline{s}\Gamma_{\mu}\frac{\lambda^a}{2}u)
(\overline{u}\Gamma^{\mu}\frac{\lambda^a}{2}b),
\end{eqnarray}
 the penguin operators are denoted by ellipses. In the above
$\Gamma_{\mu} =\gamma_{\mu}(1-\gamma_{5})$,
$Tr(\lambda^a\lambda^b)=2\delta^{ab}$.
To the operator $O_1$, we employ the results
of QCD factorization approach directly to the contributions from
the factorizaton and $\alpha_{s}$ corrections since the result of
 LCSR is
consistent with the prediction of the QCD factorization approach.

In order to calculate the nonfactorizable matrix elements
induced by the operator $\widetilde{O}_{1}$, we choose a proper
vaccum-kaon correlation function:

\begin{equation}
F_{\alpha}^{(\widetilde{O}_{1})}(p,q,k)=-\int d^4xe^{-i(p-q)x}\int
d^4ye^{i(p-k)}\langle0|
T\{j_{\alpha5}^{(K)}(y)\widetilde{O}_{1}(0)j_{5}^{(B)}(x)\}|K^-(q)\rangle,
\end{equation}
where $j_{\alpha5}^{(K)}=\overline{u}\gamma_{\alpha}\gamma_{5}s$ and
$j_{5}^{(B)}=m_{b}\overline{b}i\gamma_{5}d$ are the quark currents
interpolating $K$ and $B$ mesons, respectively.
The decomposition of the correlation
function Eq.(5) in terms of
independent momenta is straightforward and contains four invariant
amplitudes:
\begin{equation}
F_{\alpha}^{(\widetilde{O}_{1})}=(p-k)_{\alpha}F^{(\widetilde{O}_{1})}+
q_{\alpha}\tilde{F_{1}}^{(\widetilde{O}_{1})}+
k_{\alpha}\tilde{F_{2}}^{(\widetilde{O}_{1})}+
\epsilon_{\alpha\beta\lambda\rho}q^{\beta}
p^{\lambda}k^{\rho}\tilde{F_{3}}^{(\widetilde{O}_{1})}.
\end{equation}
In what follows only the amplitude $F^{(\widetilde{O}_{1})}$ is relevant.
To obtain $F^{(\widetilde{O}_{1})}$, we calculate the correlation function
by expanding the T-product of three
operators, two currents and $\widetilde{O}_{1}$, near the light-cone $x^2
\sim y^2 \sim (x-y)^2
\sim 0$. To stay away from hadronic thresholds in both channels of $K$
and $B$ currents, we choose the following kinematical region in Eq.(5):

\begin{equation}
q^2=p^2=k^2=0 \quad\mbox{and}\quad |(p-k)^2|\sim |(p-q)^2|\sim
|P^2|\gg{\Lambda_{QCD}}^2,
\end{equation}
where $P\equiv p-k-q$.

 Following the standard procedure for QCD
sum rule calculation, we can obtain:
\begin{eqnarray}
A^{(\widetilde{O}_{1})}(B\rightarrow KK)=&&\langle K(p), K(-q)
|\widetilde{O}_{1}|B(p-q)
\rangle \nonumber \\ &&
=\frac{-i}{\pi^2 f_{K}f_{B}{m_{B}}^2}\int_{0}^{s_{0}^{K}}dse^
{\frac{-s}{M^2}}\int_{m_b^2}^{\bar{R}(s,m_b^2,m_B^2,s_0^B)}
ds^{\prime} \nonumber \\ &&
e^{\frac{m_B^2-s^{\prime}}{{M^{\prime}}^2}}Im_{s^{\prime}}Im_{s}
F_{QCD}^{(\widetilde{O}_{1})}(s,s^{\prime},m_B^2),
\end{eqnarray}
where $s_{0}^{K}$ and $s_{0}^{B}$ are effective threshold parameters.
A straightforward calculation gives the following results for the
twist-3 and twist-4 contributions:

\begin{equation}
F^{(\widetilde{O}_{1})}_{QCD}=F_{tw3}^{(\widetilde{O}_{1})}+
F_{tw4}^{(\widetilde{O}_{1})},
\end{equation}
with
\begin{eqnarray}
F_{tw3}^{(\widetilde{O}_{1})}=&&\frac{m_b f_{3{K}}}{4
{\pi}^2}\int_{0}^{1}
dv \int D\alpha_{i} \nonumber  \\ &&
\times \frac{\varphi_{3K}(\alpha_{i})}{(m_b^2-(p-q)^2(1-\alpha_{1}))
(-P^2v \alpha_{3}-(p-k)^2(1-v \alpha_{3}))} \nonumber  \\ &&
\times[(2-v)(q \cdot k)+2(1-v)q \cdot (p-k)](q \cdot (p-k)),
\end{eqnarray}
and

\begin{eqnarray}
F_{tw4}^{(\widetilde{O}_{1})}=&&-\frac{m_b^2f_{K}}{4 {\pi}^2}
\int_{0}^{1}dv \int
D\alpha_{i}\widetilde{\varphi}_{\perp}(\alpha_{i})\frac
{1}{m_b^2-(p-q+q \alpha_{1})^2}\frac
{(4v-6)(p-k)q}{(p-k-qv\alpha_{3})^2} \nonumber \\ &&
+\frac{m_b^2f_{K}}{2 {\pi}^2}\int_{0}^{1}dv \int d\alpha_{1}
d\alpha_{3} \Phi_{1}(\alpha_{1},\alpha_{3})\frac
{1}{[m_b^2-(p-q+q \alpha_{1})^2]^2}\frac{(2pq-2vqk)(p-k)q}
{(p-k-qv\alpha_{3})^2} \nonumber \\&&
-\frac{m_b^2f_{K}}{2 {\pi}^2}\int_{0}^{1}dv \int d\alpha_{3} \Phi_{2}
(\alpha_{3})\frac {1}{[m_b^2-(p-q\alpha_{3})^2]^2}\frac{(2pq-2vqk)(p-k)q}
{(p-k-qv\alpha_{3})^2} \nonumber \\&&
+\frac{m_b^2f_{K}}{2 {\pi}^2}\int_{0}^{1}dv2v^2\int d\alpha_{3} \Phi_{2}
(\alpha_{3})\frac {1}{pq[m_b^2-(p-q\alpha_{3})^2]}\frac{[(p-k)q]^3}
{(p-k-qv\alpha_{3})^4} \nonumber \\&&
-\frac{m_b^2f_{K}}{2 {\pi}^2}\int_{0}^{1}dv(2v-2)v\int d\alpha_{3}
\Phi_{2}(\alpha_{3})
\frac {1}{m_b^2-(p-q\alpha_{3})^2}\frac{[(p-k)q]^2}
{(p-k-qv\alpha_{3})^4},
\end{eqnarray}
where the definitions of $\varphi_{3K}(\alpha_{i})$,
$\varphi_{\parallel}(\alpha_{i})$ and
$\varphi_{\perp}(\alpha_{i})$ can be found in Ref.\cite{s5}.

By taking the duality approximation and applying Borel
 transformation, we get the following sum rule:
\begin{eqnarray}
A^{(\widetilde{O}_{1})}&&(B\rightarrow KK) \nonumber \\&&
=im_B^2
(\frac{1}{4{\pi}^2 f_{K}} \int_{0}^{s_{0}^{K}}dse^{-\frac{s}{M^2}})
(\frac{m_b^2}{2f_B m_B^4}\int_{u_0^B}^{1} \frac{du}{u}
e^{\frac{m_B^2}{{M^{\prime}}^2}- \frac{m_b^2}
{u{M^{\prime}}^2}} \nonumber \\&&
\times[\frac{m_bf_{3K}}{u}\int_{0}^{u}\frac{dv}{v}\varphi_{3K}
(1-u,u-v,v)+f_{K}\int_{0}^{u}\frac{dv}{v}[3 \widetilde{\varphi}_{\perp}
(1-u,u-v,v) \nonumber \\&&
-(\frac{m_b^2}{u{M^{\prime}}^2}-1)\frac{\Phi_{1}(1-u,v)}
{u}]+f_{K}(\frac{m_b^2}{u{M^{\prime}}^2}-2)\frac{\Phi_{2}(u)}{u^2}]),
\end{eqnarray}
where the light-cone wave functions are introduced as the following:
\begin{eqnarray}
&&\frac{\partial {\Phi_{1}}(w,v)}{\partial v
}=\widetilde{\varphi}_{\perp}
(w,1-w-v,v)+\widetilde{\varphi}_{\parallel}(w,1-w-v,v) \nonumber \\&&
\frac{\partial {\Phi_{2}}(v)}{\partial v}=\Phi_{1}(1-v,v),
\end{eqnarray}
For the kaon distribution amplitudes in Eq.(12), we employ
their asymptotic forms which were given by Ref.[5]:
\begin{eqnarray}
&&\varphi_{3K}(\alpha_i)=360 \alpha_{1}  \alpha_{2} \alpha_{3}^2,
\quad
\widetilde{\varphi}_{\perp}(\alpha_i)=10\delta^2\alpha_{3}^2
(1-\alpha_{3}), \nonumber \\&&
\widetilde{\varphi}_{\parallel}(\alpha_{i})=-40\delta^2\alpha_{1}
\alpha_{2}\alpha_{3}.
\end{eqnarray}

\section{DECAY AMPLITUDES WITH THE SOFT-GLUON CORRECTIONS}

The hadronic matrix elements of the penguin operator
can be obtained from the same procedure. In fact,
they can be represented by the
tree diagram operator's exactly. So the calculations are simplified
relatively. Here, to compare with
the calculated results, we write
down the $B \rightarrow KK$ decay amplitudes for all decay channels
in terms of the sum of three parts:
 the factorization part $M_f$, the $\alpha_s$
correction term $M_{\alpha_s}$ and the soft-gluon contribution
$M_{nf}$:
\begin{eqnarray}
M_{f+\alpha_{s}}(\bar{B}_{s}^0 \rightarrow K^{0} \bar{K}^{0})
=&&-i\frac{G_{F}}{\sqrt{2}}f_{K}F_{0}^{B\rightarrow K}
(0)(m_B^2-m_{K}^2) \nonumber \\&&
\times
\{V_{tb}V_{ts}^*[a_{4}-\frac{1}{2}a_{10}+(a_{8}-2a_{6})R_{1}]\},
\end{eqnarray}

\begin{equation}
M_{nf}(\bar{B}_{s}^0 \rightarrow K^{0} \bar{K}^{0})
=-\frac{G_{F}}{\sqrt{2}}[V_{tb}V_{ts}^*(2c_{3}-c_{9})]
A^{(\widetilde{O}_{1})},
\end{equation}
with $R_{1}=\frac{m_{K}^2}{(m_s+m_d)(m_d-m_b)}$.
\begin{eqnarray}
M_{f+\alpha_{s}}(\bar{B}_{s}^0 \rightarrow K^{+}K^{-})
=&&i\frac{G_{F}}{\sqrt{2}}f_{K}F_{0}^{B\rightarrow K}
(0)(m_B^2-m_{K}^2)\{V_{ub}V_{us}^*a_{1} \nonumber \\&&
-V_{tb}V_{ts}^*[a_4-2(a_{6}+a_{8})R_{2}+a_{10}]\}
\end{eqnarray}

\begin{equation}
M_{nf}(\bar{B}_{s}^0 \rightarrow K^{+}K^{-})
=\sqrt{2}G_{F}V_{ub}V_{us}^*c_{2}A^{(\widetilde{O}_{1})}
-\sqrt{2}G_{F}V_{tb}V_{ts}^*
(c_{3}+c_{9})A^{(\widetilde{O}_{1})},
\end{equation}
with $R_{2}=\frac{m_{K}^2}{(m_u-m_b)(m_s+m_u)}$.
\begin{eqnarray}
M_{f+\alpha_{s}}(B^{-} \rightarrow K^{-}K^{0})
=&&-i\frac{G_{F}}{\sqrt{2}}f_{K}F_{0}^{B\rightarrow K}(0)
(m_B^2-m_{K}^2) \nonumber \\&&
\times \{V_{tb}V_{td}^*
[a_{4}-\frac{1}{2}a_{10}+(-2a_{6}+a_{8})R_{3}]\},
\end{eqnarray}

\begin{eqnarray}
M_{nf}(B^{-} \rightarrow K^{-}
K^{0})=-\frac{G_{F}}{\sqrt{2}}[V_{tb}V_{td}^*
(2c_{3}-c_{9})]A^{(\widetilde{O}_{1})}.
\end{eqnarray}
with $R_{3}=\frac{m_{K}^2}{(m_s-m_b)(m_s+m_u)}$.
In order to do the numerical calculation, we take \cite{s1} $f_{K}=160MeV,
s_0^{K}=1.62GeV^2$ and $M^2=0.5-1.2GeV^2$ for the parameters of
the kaon channel. For the B meson, we put $f_B=180 MeV$, $m_b=4.7GeV$,
$s_0^{B}=35 GeV^2$,
$\mu_{b}=\sqrt {m_B^2-m_b^2} \approx 2.4 GeV$,
${M^{\prime}}^2=8-12 GeV^2$, $f_{3K}(\mu_{b})=0.0035 GeV^2$,
${\delta}^2(\mu_b)=0.17 GeV^2$, $f_{BK}^{+}=0.32$ \cite{s6},
and $f_{B_{s}K}^{+}=0.27$ \cite{s7}. The
values of
Wilson coefficients $c_{i}$, coefficients $a_{i}$, and scale
parameter
$\mu=\frac{m_b}{2}$ are taken from Ref. \cite{s8}.

Focusing on a numerical comparison of $M_f$, $M_{\alpha_s}$ and
$M_{nf}$, we write down the numerical results for the decay
amplitudes $M=M^T+M^P$ by defining the tree amplitudes $M^T=
M_f^T+M^T_{\alpha_s}+M^T_{nf}$ and the penguin amplitudes $M^P=
M_f^P+M^P_{\alpha_s}+M^P_{nf}$:
\begin{eqnarray}
&&M(\bar{B}_{s}^{0} \rightarrow K^{0} \bar{K}^{0})=
M^P(\bar{B}_{s}^{0} \rightarrow K^{0} \bar{K}^{0})\nonumber\\
&=&V_{tb}V_{ts}^*\{[8.34568 \times 10^{-7}i]+[-2.96515 \times
10^{-7}-1.39319 \times 10^{-7}i]+[-1.08188 \times 10^{-8}i]\},
\nonumber\\
\end{eqnarray}
\begin{eqnarray}
&&M(\bar{B}_{s}^{0} \rightarrow K^{+} K^{-})=
M^T(\bar{B}_{s}^{0} \rightarrow K^{+} K^{-})+
M^P(\bar{B}_{s}^{0} \rightarrow K^{+} K^{-})\nonumber\\
&=&V_{ub}V_{us}^*\{[1.02213 \times 10^{-5}i]+[-3.24748 \times
10^{-7}+3.77232 \times 10^{-7}i]+[-1.22115 \times
10^{-7}i]\}\nonumber\\
&+&V_{tb}V_{ts}^*\{[8.4483 \times
10^{-7}i]+[-2.97568 \times 10^{-7}-1.56528 \times 10^{-7}i]+
[-4.44113
\times 10^{-9}i]\},\nonumber\\
\end{eqnarray}
\begin{eqnarray}
&&M(B^{-} \rightarrow K^{0} K^{-})=
M^P(B^{-} \rightarrow K^{0} K^{-})\nonumber\\
&=&V_{tb}V_{td}^*\{[1.02012 \times 10^{-6}i]+[-3.59909 \times
10^{-7}-1.68011 \times 10^{-7}i]+[-1.08188 \times 10^{-8}i]\}.
\nonumber\\
\end{eqnarray}

 \begin{figure}[ht]
 \includegraphics[200,155][400,485]{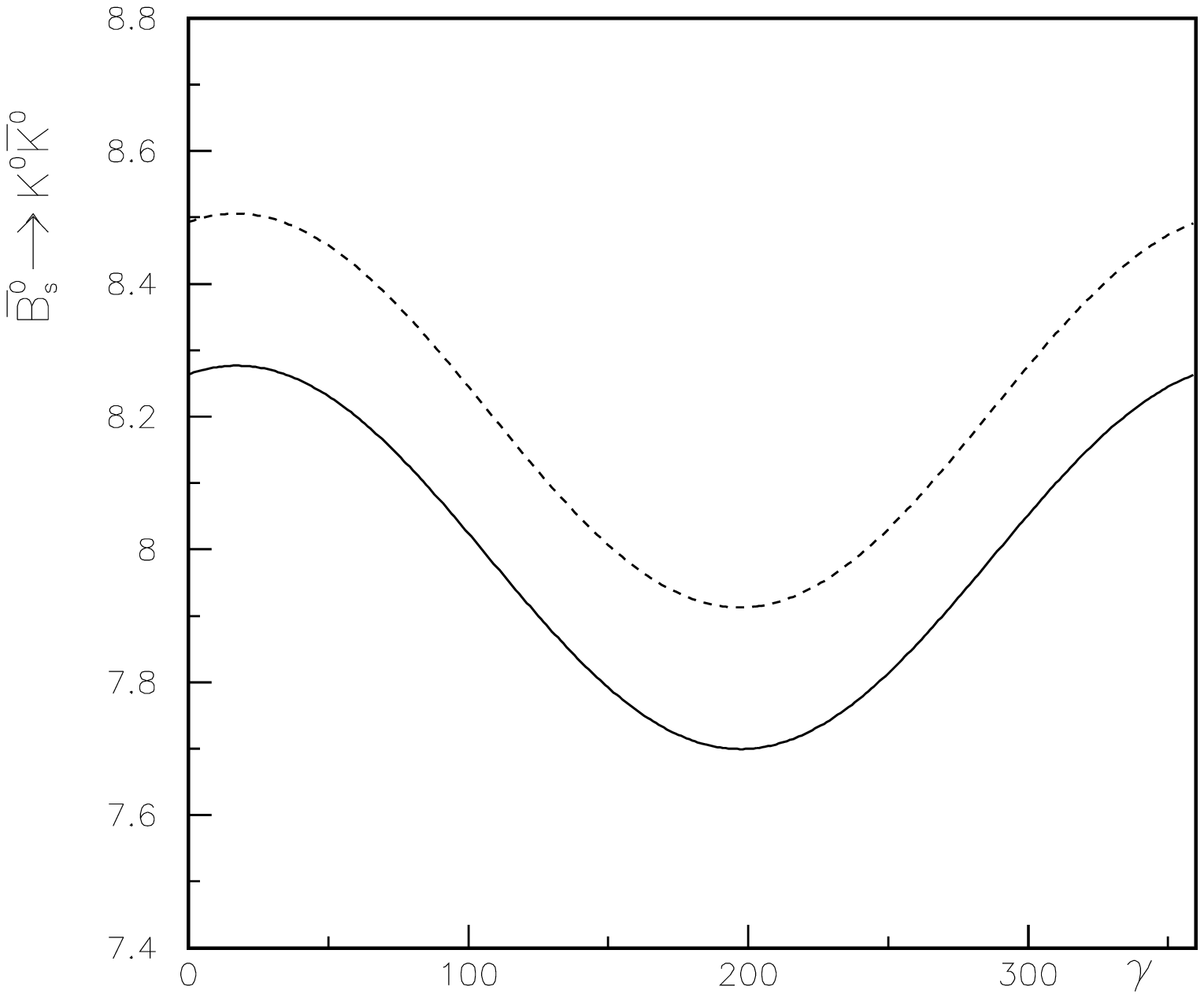}
 \caption{Dependence of the branching ratio on the weak phase $\gamma$
  in the $\bar{B_{s}}^{0} \rightarrow K^{0}\bar{K}^{0}$
  channel. The dashed and solid lines correspond
  to the values obtained with
  $M_f+M_{\alpha_s}$ and $M_f+M_{\alpha_s}+M_{nf}$,
  respectively.}
 \label{fig1}
 \end{figure}
  \begin{figure}[ht]
  \includegraphics[200,155][400,485]{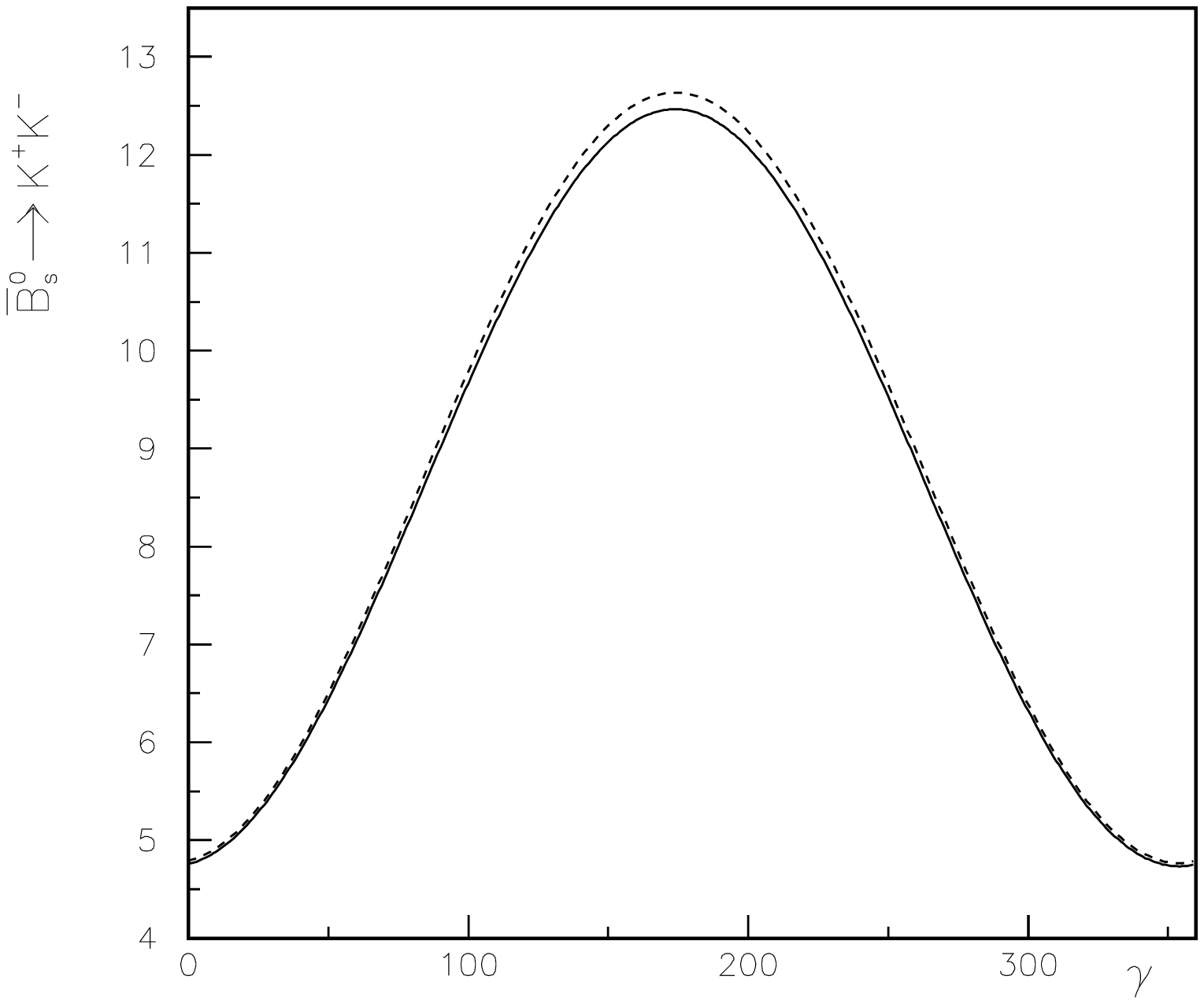}
  \caption{Dependence of the branching ratio on the weak phase $\gamma$
   in the $\bar{B_{s}}^{0} \rightarrow K^{+}K^{-}$
   channel. The dashed and solid lines correspond
   to the values obtained with
   $M_f+M_{\alpha_s}$ and $M_f+M_{\alpha_s}+M_{nf}$,
   respectively.}
   \label{fig2}
  \end{figure}
  \begin{figure}[ht]
  \includegraphics[200,155][400,485]{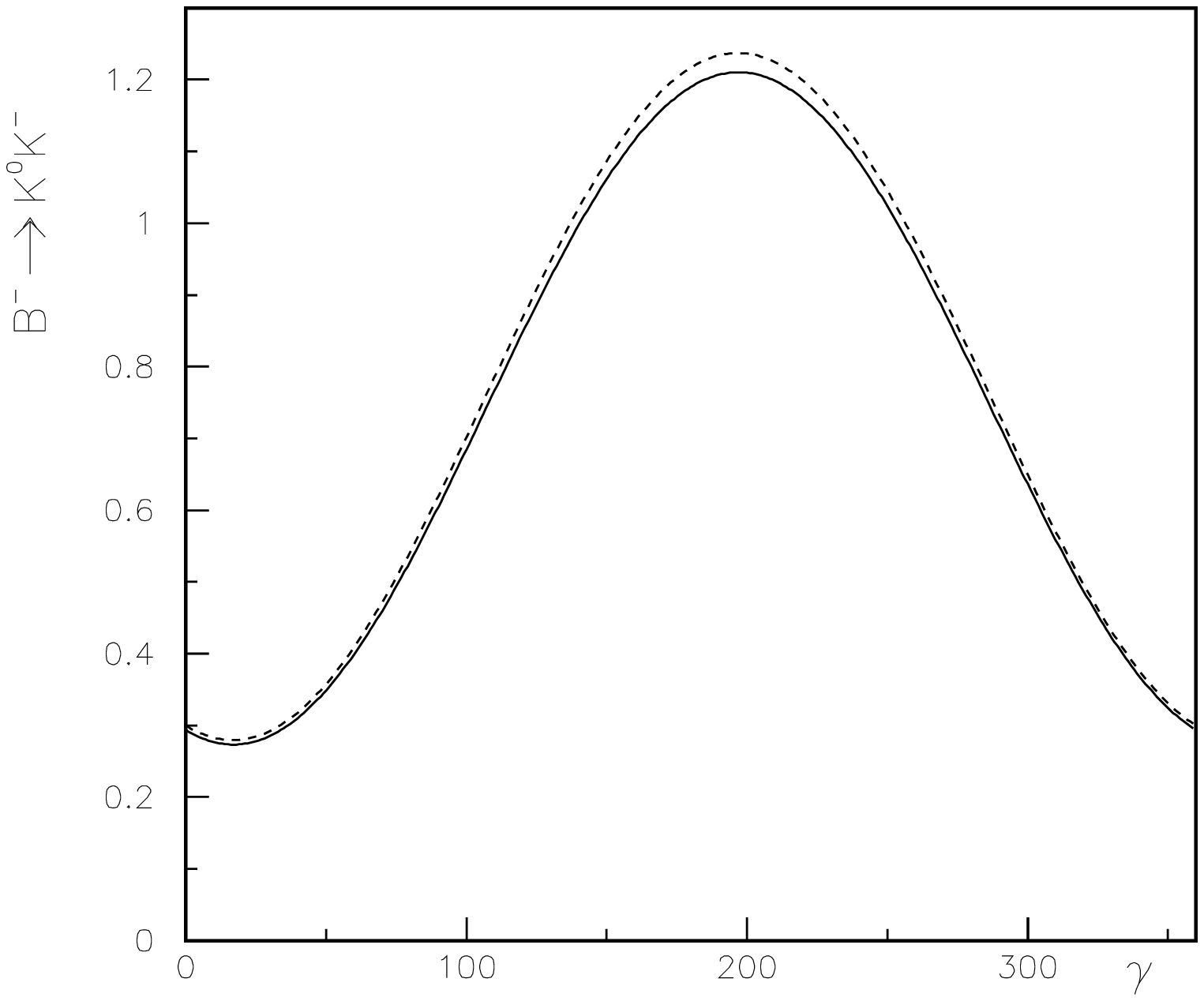}
  \caption{Dependence of the branching ratio on the weak phase $\gamma$
   in the $B^{-} \rightarrow K^{0}K^{-}$
    channel. The dashed and solid lines correspond
    to the values obtained with
    $M_f+M_{\alpha_s}$ and $M_f+M_{\alpha_s}+M_{nf}$,
   respectively.}
   \label{fig3}
    \end{figure}

For comparison, we plot the branching
ratios (Br) for these decay modes as a function of $\gamma$
in Fig.1-Fig.3.
Eqs.(21)-(23) show that the soft-gluon contributions to
the decay amplitudes are much smaller than the
factorization
parts in $B \rightarrow KK$ decays. The contributions
depend on different decay modes. In the case
of
$\bar{B_{s}}^{0} \rightarrow K^{0}\bar{K}^{0}$, there
are only
penguin diagram contributions and they come mainly from the
factorization and $ \alpha_{s}$ correction parts; soft-gluon
contribution is suppressed by the order of $10^{-1}$ with
respect to the former. The soft-gluon effects make the
branching ratio smaller and the results are shown in Fig.1.
The dashed and solid curves correspond to the values obtained
from $M_f+M_{\alpha_s}$ and $M_f+M_{\alpha_s}+M_{nf}$,
respectively. In the case of
$\bar{B_{s}}^{0} \rightarrow K^{+}K^{-}$, the soft
contribution has the same order as the $\alpha_{s}$ correction
 parts in the tree amplitudes. In the penguin amplitudes, it
has the amplitude of order $10^{-9}$, which is
smaller than those of factorization and $\alpha_{s}$
correction parts (of order $10^{-7}$).
It is shown from Fig.2 that the
total branching ratio is suppressed by soft-gluon
 effects. In the case of
$B^{-} \rightarrow K^{0}K^{-}$, the soft-gluon amplitude
is of order $10^{-8}$, which is obviously lower than that of
the factorization (of order $10^{-6}$) and $\alpha_{s}$
correction (of order $10^{-7}$). Fig.3 show that
the total branching ratio is decreased by soft-gluon
effects, too.

\section{SUMMARY}

In this paper, we have analyzed the $B \rightarrow KK$ decays
with the soft-gluon corrections by the QCD light-cone sum
rules. The soft contributions
depend on decay modes, and in most situations they have
the amplitudes which are suppressed by the order of $10^{-1}$
or $10^{-2}$ of factorization and $\alpha_{s}$ correction
amplitudes. Only in the case of
$\bar{B}_{s}^{0} \rightarrow K^{+} K^{-}$,
they have the same
order amplitude with $\alpha_{s}$ correction parts
which is smaller than the factorization amplitude.
 Our results show that the soft-gluon
 effects on $B \rightarrow KK$ are small
and they always suppress the branching ratio values with
 2-3 $\%$ corrections .

\end{document}